\documentstyle[12pt]{article}
\begin{document}
\large
\bibliographystyle{plain}
\begin{titlepage}
\large
\hfill\begin{tabular}{l}HEPHY-PUB 632/96\\ UWThPh-1995-23\\ March 1996
\end{tabular}\\[2.5cm]
\begin{center}
{\Large\bf RELATIVISTIC COULOMB PROBLEM:}\\[.5ex]
{\Large\bf ANALYTIC UPPER BOUNDS}\\[.5ex]
{\Large\bf ON ENERGY LEVELS}\\
\vspace{1.8cm}
{\Large\bf Wolfgang LUCHA}\\[.5cm]
Institut f\"ur Hochenergiephysik,\\
\"Osterreichische Akademie der Wissenschaften,\\
Nikolsdorfergasse 18, A-1050 Wien, Austria\\[1.5cm]
{\Large\bf Franz F.~SCH\"OBERL}\\[.5cm]
Institut f\"ur Theoretische Physik,\\
Universit\"at Wien,\\
Boltzmanngasse 5, A-1090 Wien, Austria\\[2cm]
{\bf Abstract}
\end{center}
\normalsize
\noindent
The spinless relativistic Coulomb problem is the bound-state problem for the
spinless Salpeter equation (a standard approximation to the Bethe--Salpeter
formalism as well as the most simple generalization of the nonrelativistic
Schr\"odinger formalism towards incorporation of relativistic effects) with
the Coulomb interaction potential (the static limit of the exchange of some
massless bosons, as present in unbroken gauge theories). The nonlocal nature
of the Hamiltonian encountered here, however, renders extremely difficult to
obtain rigorous analytic statements on the corresponding solutions. In view
of this rather unsatisfactory state of affairs, we derive (sets of) analytic
upper bounds on the involved energy eigenvalues.\\

\large
\end{titlepage}

\section{Introduction: The Spinless Salpeter Equation}

Maybe one of the most straightforward generalizations of the standard
nonrelativistic quantum theory towards the reconciliation with all the
requirements imposed by special relativity is represented by describing the
quantum systems under consideration by the well-known ``spinless Salpeter
equation.'' Consider a quantum system the dynamics of which is governed by a
by assumption self-adjoint Hamiltonian $H$ of the form
\begin{equation}
H = T + V\ ,
\label{eq:ham-sseq}
\end{equation}
where $T$ denotes the square-root operator of the relativistic expression for
the free (kinetic) energy of some particle of mass $m$ and momentum ${\bf
p}$,
$$
T \equiv \sqrt{{\bf p}^2 + m^2}\ ,
$$
and $V = V({\bf x})$ represents some arbitrary coordinate-dependent static
interaction potential. The eigenvalue equation for this Hamiltonian $H$,
$$
H|\chi_k\rangle = E_k|\chi_k\rangle\ ,\quad k = 0,1,2,\dots\ ,
$$
for Hilbert-space eigenvectors $|\chi_k\rangle$ corresponding to energy
eigenvalues
$$
E_k \equiv
\frac{\langle\chi_k|H|\chi_k\rangle}{\langle\chi_k|\chi_k\rangle}\ ,
$$
is nothing else but the one-particle spinless Salpeter equation. Because of
the nonlocality of this operator $H$, that is, more precisely, of either the
kinetic-energy operator $T$ in configuration space or the
interaction-potential operator $V$ in momentum space, it is hard to obtain
analytic statements from this equation of motion. (For the ``translation'' of
the equal-mass two-particle problem to the one-particle problem discussed at
present, see Appendix~\ref{app:1-2-part-ham}.)

\section{The Spinless Relativistic Coulomb Problem}

Of particular importance in all physics are the (spherically symmetric)
central potentials, which depend only on the radial coordinate $r \equiv
|{\bf x}|$. Among these, the most prominent one is the Coulomb potential
$V_{\rm C}(r)$, which is parametrized by some (dimensionless) coupling
constant $\alpha$:
\begin{equation}
V({\bf x}) = V_{\rm C}(r) = - \frac{\alpha}{r} \ , \quad \alpha > 0 \ .
\label{eq:coulpot}
\end{equation}
The (semi-)relativistic Hamiltonian (\ref{eq:ham-sseq}) with the Coulomb
interaction potential $V_{\rm C}$ in (\ref{eq:coulpot}) defines the
``spinless relativistic Coulomb problem.'' In the past, the interest in this
spinless relativistic Coulomb problem has undergone an eventful history. (For
a rather comprehensive review, consult Ref.~\cite{lucha94}.) Let us merely
sketch in the following some highlights.

First of all, from an examination \cite{herbst77} of the spectral properties
of the operator (\ref{eq:ham-sseq}), (\ref{eq:coulpot}) one may infer the
existence of its Friedrichs extension up to the critical value
$$
\alpha_{\rm c} = \frac{2}{\pi}
$$
of the involved coupling constant $\alpha$ and read off a lower bound on the
corresponding ground-state energy eigenvalue $E_0$, namely,
\begin{equation}
E_0 \ge m\,\sqrt{1 - \left(\frac{\pi\,\alpha}{2}\right)^2}\quad\mbox{for}\
\alpha < \frac{2}{\pi}\ ,
\label{eq:lb-herbst}
\end{equation}
which, later on, has been slightly improved to \cite{martin89}
$$
E_0 \ge m\,\sqrt{\frac{1 + \sqrt{1 - 4\,\alpha^2}}{2}}\quad\mbox{for}\
\alpha < \frac{1}{2}\ .
$$
The analytic solutions for the wave functions of those eigenstates
$|\chi\rangle$ of the Hamiltonian (\ref{eq:ham-sseq}), (\ref{eq:coulpot})
which correspond to vanishing orbital angular momentum have been constructed
\cite{durand83}. The attempt in \cite{durand83} to determine simultaneously
the respective (set of) exact energy eigenvalues without actually solving
this spinless Salpeter equation, however, failed
\cite{hardekopf85,lucha94varbound}. Therefore, as far as analytic statements
about the relativistic Coulomb problem, in particular, its energy
eigenvalues, are concerned, up to now one has to content oneself with a few
series expansions of these energy eigenvalues $E_k$ in powers of the coupling
constant $\alpha$ \cite{leyaouanc94,brambilla95}, which then are, of course,
only significant for a region of rather small values of $\alpha$.

\section{Analytic Upper Bounds on Energy Eigenvalues}

Without a closed form of all the energy eigenvalues $E_k$ of the spinless
relativistic Coulomb problem at hand, it is highly desirable to have, at
least, analytic expressions for upper bounds on these at one's disposal, in
order to estimate the reliability of approximative solutions or series
expansions.

The theoretical basis as well as the primary tool for the derivation of
rigorous upper bounds on the eigenvalues of some self-adjoint operator is,
beyond doubt, the so-called ``min--max principle'' \cite{reed-simon}. An
immediate consequence of this min--max principle is the following statement:
Let $H$ be a self-adjoint operator that is bounded from below [as, according
to Eq.~(\ref{eq:lb-herbst}), evidently holds for the (semi-)relativistic
Hamiltonian in (\ref{eq:ham-sseq}) with a Coulomb-type interaction potential
(\ref{eq:coulpot})]. Let $E_k$, $k = 0,1,2,\dots$, denote the eigenvalues of
$H$, ordered according to $E_0 \le E_1 \le E_2 \le \dots$. Let $D_d$ be some
$d$-dimensional subspace of the domain of $H$. Then the $k$th eigenvalue
(counting multiplicity) of $H$, $E_k$, satisfies the inequality
\begin{equation}
E_k \le \sup_{\psi \in D_{k+1}}
\frac{\langle\psi|H|\psi\rangle}{\langle\psi|\psi\rangle}\ ,\quad
k = 0,1,2,\dots\ .
\label{eq:min-max}
\end{equation}

A peculiarity of the (spinless relativistic) Coulomb problem is that there is
only one dimensional parameter, namely, the particle mass $m$. As a
consequence of this, for a vanishing particle mass, i.~e., for $m=0$, the
totality of eigenvalues $E_k$ of the Hamiltonian (\ref{eq:ham-sseq}) with
interaction potential (\ref{eq:coulpot}) collapses, already on dimensional
grounds, necessarily to $E_k = 0$ for all $k=0,1,2,\dots$. This fact is also
clearly demonstrated by application of the ``relativistic virial theorem''
proven in Refs.~\cite{lucha89rvt,lucha90rvt}. Accordingly, in the course of
an investigation of the spinless relativistic Coulomb problem, it is
sufficient to focus one's interests to the special case of a nonvanishing
particle mass $m$:
$$
m > 0\ .
$$

In view of the above, when searching for bounds, our intention must be to
avoid in some way or other the problematic square-root operator in order to
deal with more manageable Hamiltonians.

\subsection{The ``Schr\"odinger'' bound}\label{sec:schrbound}

From the positivity of the square of the obviously self-adjoint operator $T -
m$,
\begin{eqnarray*}
0 &\le& (T - m)^2\\[1ex]
&=& T^2 + m^2 - 2\,m\,T\\[1ex]
&=& {\bf p}^2 + 2\,m^2 - 2\,m\,T\ ,
\end{eqnarray*}
one obtains, for the free (or kinetic) energy $T$, the operator inequality
$$
T \le m + \frac{{\bf p}^2}{2\,m}\ ,
$$
and thus, for the generic Hamiltonian $H$ in (\ref{eq:ham-sseq}), the
operator inequality
\begin{equation}
H \le H_{\rm S}\ ,
\label{eq:opineq}
\end{equation}
where $H_{\rm S}$ denotes the ``Schr\"odinger'' Hamiltonian
\begin{equation}
H_{\rm S} = m + \frac{{\bf p}^2}{2\,m} + V\ .
\label{eq:ham-schr}
\end{equation}
Applying to the energy eigenvalues $E_k$ of the Hamiltonian $H$ in
Eq.~(\ref{eq:ham-sseq}) first the min--max principle in the form given by
Eq.~(\ref{eq:min-max}) and after that the operator inequality
(\ref{eq:opineq}), we find
\begin{eqnarray*}
E_k &\equiv&
\frac{\langle\chi_k|H|\chi_k\rangle}{\langle\chi_k|\chi_k\rangle}\\[1ex]
&\le& \sup_{\psi \in D_{k+1}}
\frac{\langle\psi|H|\psi\rangle}{\langle\psi|\psi\rangle}\\[1ex]
&\le& \sup_{\psi \in D_{k+1}}
\frac{\langle\psi|H_{\rm S}|\psi\rangle}{\langle\psi|\psi\rangle}\ .
\end{eqnarray*}
Now, let us assume that the $(k+1)$-dimensional subspace $D_{k+1}$ in this
inequality is spanned by the first $k+1$ eigenvectors of the Schr\"odinger
Hamiltonian $H_{\rm S}$, that is, by precisely those eigenvectors of $H_{\rm
S}$ which correspond to the first $k+1$ energy eigenvalues $E_{{\rm S},0},
E_{{\rm S},1},\dots,E_{{\rm S},k}$ when all eigenvalues of $H_{\rm S}$ are
ordered according to $E_{{\rm S},0}\le E_{{\rm S},1}\le E_{{\rm
S},2}\le\dots$. In this case, the right-hand side of the above inequality is
nothing else but the $k$th ``Schr\"odinger'' energy eigenvalue $E_{{\rm
S},k}$:
$$
\sup_{\psi \in D_{k+1}}
\frac{\langle\psi|H_{\rm S}|\psi\rangle}{\langle\psi|\psi\rangle}
= E_{{\rm S},k}\ .
$$
Consequently, any single energy eigenvalue $E_k$ of the spinless Salpeter
equation is bounded from above by its ``Schr\"odinger'' counterpart $E_{{\rm
S},k}$:
\begin{equation}
E_k \le E_{{\rm S},k}\ .
\label{eq:schrub}
\end{equation}

For the Coulomb potential (\ref{eq:coulpot}), the energy eigenvalues required
here are well known:
\begin{equation}
E_{{\rm S},n} = m\left(1 - \frac{\alpha^2}{2\,n^2}\right)\ ,
\label{eq:schrenevs}
\end{equation}
where the total quantum number $n$ is given in terms of both radial and
orbital angular momentum quantum numbers $n_{\rm r}$ and $\ell$,
respectively, by
$$
n = n_{\rm r} + \ell + 1\ ,
\quad n_{\rm r} = 0,1,2,\dots\ , \quad \ell = 0,1,2,\dots\ .
$$

\subsection{A ``squared'' bound}

One might be tempted to try to find an improvement of the bound
(\ref{eq:schrub}) by considering the square of the Hamiltonian $H$, that is,
the operator
$$
Q \equiv H^2 = T^2 + V^2 + T\,V + V\,T\ .
$$
The eigenvalue equation for this squared Hamiltonian $Q$ will, of course, be
solved by the same set of eigenvectors $|\chi_k\rangle$ as the one for the
original Hamiltonian $H$ with, however, the squares of the corresponding
energy eigenvalues $E_k$ as the eigenvalues of $Q$:
$$
Q|\chi_k\rangle = E_k^2|\chi_k\rangle\ ,\quad k = 0,1,2,\dots\ .
$$
From the positivity of the square of the self-adjoint operator $T - m - V$,
\begin{eqnarray*}
0 &\le& (T - m - V)^2\\[1ex]
&=& T^2 + m^2 + V^2 - 2\,m\,T + 2\,m\,V - T\,V - V\,T\ ,
\end{eqnarray*}
we obtain, with the help of the obvious relation
$$
0 \le m \le T\ ,
$$
for the anticommutator $T\,V + V\,T$ of kinetic energy $T$ and interaction
potential $V$, which appears in the square $Q$ of our Hamiltonian $H$, the
operator inequality
\begin{eqnarray*}
T\,V + V\,T
&\le& T^2 + m^2 + V^2 - 2\,m\,T + 2\,m\,V\\[1ex]
&\equiv& {\bf p}^2 + 2\,m^2 + V^2 - 2\,m\,T + 2\,m\,V\\[1ex]
&\le& {\bf p}^2 + V^2 + 2\,m\,V\ ,
\end{eqnarray*}
which, when applied to the squared Hamiltonian $Q$, yields the operator
inequality
$$
Q \le R\ ,
$$
with the operator $R$ given by
$$
R \equiv 2\,{\bf p}^2 + m^2 + 2\,V^2 + 2\,m\,V\ .
$$
Following the line of argument as given in the preceding subsection, we might
therefore conclude that the squares of the energy eigenvalues $E_k$ of the
spinless relativistic Coulomb problem are bounded from above by the
eigenvalues ${\cal E}_{{\rm R},k}$ of the operator $R$:
$$
E_k^2 \le {\cal E}_{{\rm R},k}\ ,
$$
which entails
\begin{equation}
E_k \le \sqrt{{\cal E}_{{\rm R},k}}\ .
\label{eq:sqbounds}
\end{equation}

Now, only for the Coulomb potential (\ref{eq:coulpot}), this latter operator
$R$ is of precisely the same structure as the Schr\"odinger Hamiltonian
(\ref{eq:ham-schr}), with, however, some kind of ``effective'' orbital
angular momentum quantum number $L$ which, for states of (genuine) orbital
angular momentum $\ell$, has to be determined from the relation
$$
L\,(L + 1) = \ell\,(\ell + 1) + \alpha^2\ ,
$$
the solution of which reads
$$
L = \frac{\sqrt{1 + 4\left[\ell\,(\ell + 1) + \alpha^2\right]} - 1}{2}\ ,\quad
\ell = 0,1,2,\dots\ .
$$
In this case, the eigenvalues ${\cal E}_{\rm R}$ of the operator $R$ may be
easily written down:
$$
{\cal E}_{{\rm R},N} = m^2\left(1 - \frac{\alpha^2}{2\,N^2}\right)\ ,
$$
with the ``effective'' total quantum number
$$
N = n_{\rm r} + L + 1\ ,\quad n_{\rm r} = 0,1,2,\dots\ .
$$
Unfortunately, it is easy to convince oneself that the ``square'' bounds
(\ref{eq:sqbounds}) obtained in this way lie above and are thus worse than
the previous Schr\"odinger bounds.

\subsection{A variational bound}

For all practical purposes, the most efficient manner for the min--max
principle to come into play is in form of the ``Rayleigh--Ritz variational
technique.'' Let us illustrate this fact just for the ground-state energy
eigenvalue $E_0$. For $k=0$, the relation (\ref{eq:min-max}) simplifies to
$$
E_0 \le \frac{\langle\psi|H|\psi\rangle}{\langle\psi|\psi\rangle}\ .
$$
That is, the ground-state energy $E_0$ is clearly less than or equal to any
expectation value of the considered Hamiltonian, $H$. The above upper bound
may, of course, be optimized by determining the smallest of all these
expectation values, at least in some chosen Hilbert-space sector.
Consequently, there is a simple recipe for the derivation of (sometimes
excellent!) exact upper bounds on this ground-state energy eigenvalue:
minimize the encountered expectation values of the Hamiltonian under
consideration,
$$
\frac{\langle\psi_\lambda|H|\psi_\lambda\rangle}
{\langle\psi_\lambda|\psi_\lambda\rangle}\ ,
$$
with respect to a suitably chosen set of Hilbert-space trial vectors
$|\psi_\lambda\rangle$, which are distinguished from each other by some
variational parameter $\lambda$.

We apply this prescription to our (semi-)relativistic Hamiltonian $H$.
However, in order to make life easy, we immediately take advantage of a
simple inequality for the expectation values of a self-adjoint operator, like
our kinetic energy $T$, with respect to arbitrary Hilbert-space states
$|\psi\rangle$ in the domain of this operator:
$$
\frac{|\langle\psi|T|\psi\rangle|}{\langle\psi|\psi\rangle}
\le \sqrt{\frac{\langle\psi|T^2|\psi\rangle}{\langle\psi|\psi\rangle}}\ .
$$
Employing this inequality, we are able to circumvent the (troublesome)
expectation values of the square-root operator of the kinetic energy $T$
(which, of course, may be always evaluated numerically; see
Ref.~\cite{lucha94}):
\begin{eqnarray*}
E_0 &\le& \frac{\langle\psi_\lambda|H|\psi_\lambda\rangle}
{\langle\psi_\lambda|\psi_\lambda\rangle}\\[1ex]
&=& \frac{\langle\psi_\lambda|T + V|\psi_\lambda\rangle}
{\langle\psi_\lambda|\psi_\lambda\rangle}\\[1ex]
&\le& \sqrt{\frac{\langle\psi_\lambda|T^2|\psi_\lambda\rangle}
{\langle\psi_\lambda|\psi_\lambda\rangle}}
+ \frac{\langle\psi_\lambda|V|\psi_\lambda\rangle}
{\langle\psi_\lambda|\psi_\lambda\rangle}\\[1ex]
&\equiv& \sqrt{\frac{\langle\psi_\lambda|{\bf p}^2|\psi_\lambda\rangle}
{\langle\psi_\lambda|\psi_\lambda\rangle} + m^2}
+ \frac{\langle\psi_\lambda|V|\psi_\lambda\rangle}
{\langle\psi_\lambda|\psi_\lambda\rangle}\ .
\end{eqnarray*}

For the Coulomb potential (\ref{eq:coulpot}), the most reasonable choice of
trial vectors is obviously one for which the coordinate-space representation
$\psi_\lambda({\bf x})$ of the trial vectors $|\psi_\lambda\rangle$ for
vanishing radial and orbital angular momentum quantum numbers is given by
hydrogen-like trial functions:
$$
\psi_\lambda({\bf x}) = \exp(- \lambda\,r)\ ,\quad\lambda > 0\ .
$$
For these trial functions, the computation of the expectation values in the
above inequality yields \cite{lucha94varbound}
$$
E_0 \le \sqrt{\lambda^2 + m^2} - \alpha\,\lambda\ .
$$
Determining the minimum of this latter set of upper bounds, we arrive at
\cite{lucha94varbound}
$$
E_0 \le E_{{\rm var},0}\ ,
$$
with the variational upper bound for the ground-state energy level $E_0$ of
the spinless relativistic Coulomb problem given by
\begin{equation}
E_{{\rm var},0} \equiv m\,\sqrt{1 - \alpha^2}\ .
\label{eq:varbound0}
\end{equation}
This variational bound, $E_{{\rm var},0}$, is lower and thus better than the
former Schr\"odinger bound on the ground-state energy level,
$$
E_{{\rm S},0} = m\left(1 - \frac{\alpha^2}{2}\right)\ ,
$$
which may be obtained from Eq.~(\ref{eq:schrenevs}) for $n_{\rm r} = \ell =
0$, implying thereby $n = 1$:
$$
E_{{\rm var},0} < E_{{\rm S},0}\quad\mbox{for}\ \alpha\neq 0\ .
$$
Consequently, the variational technique entails indeed improved upper bounds
on the energy levels as compared to the Schr\"odinger estimates.

\subsection{A straightforward generalization}

Our variational upper bound (\ref{eq:varbound0}) for the ground-state energy
level $E_0$ can be very easily re-derived and simultaneously extended to
arbitrary levels of excitation by a generalization of the considerations
presented in Subsection~\ref{sec:schrbound}. Introducing an arbitrary real
parameter $\mu$ (with the dimension of mass), we make use of the positivity
of the square of the obviously self-adjoint operator $T - \mu$,
\begin{eqnarray*}
0 &\le& (T - \mu)^2\\[1ex]
&=& T^2 + \mu^2 - 2\,\mu\,T\\[1ex]
&=& {\bf p}^2 + m^2 + \mu^2 - 2\,\mu\,T\ ,
\end{eqnarray*}
in order to find, for the kinetic energy $T$, a set of operator inequalities,
$$
T \le \frac{{\bf p}^2 + m^2 + \mu^2}{2\,\mu}\quad\mbox{for all}\ \mu > 0\ ,
$$
and, consequently, for the (semi-)relativistic Hamiltonian $H$ in
Eq.~(\ref{eq:ham-sseq}), the set of operator inequalities
$$
H \le \widehat H_{\rm S}(\mu)\quad\mbox{for all}\ \mu > 0\ ,
$$
with the Schr\"odinger-type Hamiltonian $\widehat H_{\rm S}(\mu)$ given by
$$
\widehat H_{\rm S}(\mu) = \frac{{\bf p}^2 + m^2 + \mu^2}{2\,\mu} + V\ .
$$
Now, mimicking the line of argument given in Subsection~\ref{sec:schrbound}
involving the min--max principle, we conclude that the set of energy
eigenvalues $E_k$, $k = 0,1,2,\dots$, of the Hamiltonian $H$, if again
ordered according to $E_0 \le E_1 \le E_2 \le \dots$, is bounded from above
by the (corresponding) set of energy eigenvalues $\widehat E_{{\rm
S},k}(\mu)$ of this Schr\"odinger Hamiltonian $\widehat H_{\rm S}(\mu)$, if
again similarly ordered according to $\widehat E_{{\rm S},0}(\mu)\le\widehat
E_{{\rm S},1}(\mu)\le\widehat E_{{\rm S},2}(\mu)\le\dots$,
$$
E_k \le \widehat E_{{\rm S},k}(\mu)\quad\mbox{for all}\ \mu > 0\ ,
$$
and, consequently, also by the minimum of all these energy eigenvalues:
$$
E_k \le \min_{\mu > 0}\widehat E_{{\rm S},k}(\mu)\ .
$$

For the Coulomb potential (\ref{eq:coulpot}), these energy eigenvalues
$\widehat E_{{\rm S},n}(\mu)$ are given by
$$
\widehat E_{{\rm S},n}(\mu) = \frac{1}{2\,\mu}
\left[m^2 + \mu^2 \left(1 - \frac{\alpha^2}{n^2}\right)\right]\ ,
$$
with precisely the same total quantum number $n$ as before. Minimizing this
latter expression with respect to the parameter $\mu$, we finally obtain
$$
\min_{\mu > 0}\widehat E_{{\rm S},n}(\mu) =
m\,\sqrt{1 - \frac{\alpha^2}{n^2}}\ .
$$
For any value of the total quantum number $n$, these bounds definitely
improve the Schr\"odinger bounds (\ref{eq:schrenevs}):
$$
\min_{\mu > 0}\widehat E_{{\rm S},n}(\mu) < E_{{\rm S},n}\quad\mbox{for}\
\alpha\neq 0\ .
$$
Clearly, for $\mu = m$, we immediately recover the Schr\"odinger approach of
Subsection~\ref{sec:schrbound}.

\section{Summary}

Horrified by the insight that the discrete spectrum of the Hamiltonian
consisting of just the relativistic kinetic energy and the static Coulomb
interaction potential is still not known exactly, we derived by different but
elementary methods at least a few complete sets of analytic upper bounds on
the respective energy eigenvalues. In every individual case, the basic idea
was to derive an operator inequality for this Hamiltonian which guarantees
that the expectation values of this Hamiltonian are bounded from above by the
expectation values of some other operator which no longer involves the
(problematic) square-root operator of the relativistic kinetic energy, and to
``embed'' this operator inequality into the well-known min--max principle for
the eigenvalues of a self-adjoint operator bounded from below. It should be
no great surprise that just that method which employs some variational
procedure yields the best of these bounds.\vspace{3ex}

\noindent
{\bf Acknowledgements}\\

We would like to thank B.~Baumgartner, H.~Grosse, H.~Narnhofer, W.~Thirring,
and J.~Yngvason for a lot of helpful discussions.

\appendix

\section{Equivalence of One-Particle and (Equal-Mass) Two-Particle
Scenarios}\label{app:1-2-part-ham}

The two relativistic Coulombic Hamiltonians for one- and two-particle
problem,
$$
H^{(1)} = \sqrt{{\bf p}^2 + m^2} - \frac{\alpha}{r}\ ,\quad
r \equiv |{\bf x}|\ ,
$$
and
$$
H^{(2)} = 2\,\sqrt{{\bf P}^2 + M^2} - \frac{\kappa}{R}\ ,\quad
R \equiv |{\bf X}|\ ,
$$
respectively, may be equated with the help of a scale transformation as
follows. Relate the employed phase-space variables, $({\bf x},{\bf p})$ and
$({\bf X},{\bf P})$, respectively, by some (in general arbitrary) scale
factor $\lambda$ according to
\begin{eqnarray*}
{\bf p} &=& \lambda\,{\bf P}\ ,\\[1ex]
{\bf x} &=& \frac{{\bf X}}{\lambda}\ ,
\end{eqnarray*}
which preserves their fundamental commutation relations:
$$
[{\bf x},{\bf p}] = [{\bf X},{\bf P}]\ .
$$
Fix this scale factor, $\lambda$, to the particular value $\lambda = 2$ and
identify both the mass and the Coulomb coupling strength parameters according
to
\begin{eqnarray*}
m &=& 2\,M\ ,\\[1ex]
\alpha &=& \frac{\kappa}{2}\ .
\end{eqnarray*}
You will end up with
$$
H^{(1)} = H^{(2)}\ .
$$

\end{document}